\documentclass[twocolumn,showpacs,amsmath,amssymb]{revtex4}

\usepackage{graphicx}
\usepackage{dcolumn}
\usepackage{bm}

\begin{document}

\preprint{APS/123-QED}
\title{Phase correlation of laser waves with arbitrary frequency spacing}
\author{A. F. Huss, R. Lammegger, C. Neureiter, E. A. Korsunsky and L. Windholz}
\affiliation{Institut f\"{u}r Experimentalphysik, Technische Universit\"{a}t Graz,
Petersgasse 16, 8010 Graz, \"{O}sterreich}
\date{March 10, 2004}

\begin{abstract}
The theoretically predicted correlation of laser phase fluctuations in $%
\Lambda$ type interaction schemes is experimentally demonstrated.
We show, that the mechanism of correlation in a $\Lambda$ scheme
is restricted to high frequency noise components, whereas in a
double-$\Lambda$ scheme, due to the laser phase locking in
closed-loop interaction, it extends to all noise frequencies. In
this case the correlation is weakly sensitive to coherence losses. Thus the double-$%
\Lambda$ scheme can be used to correlate e.m. fields with carrier frequency
differences beyond the GHz regime.
\end{abstract}

\maketitle

The study of quantum interference effects in optical dense media,
such as Electromagnetically Induced Transparency (EIT)
\cite{harPT50}, is one of the most challenging fields in modern
quantum optics research. In ideal EIT atoms are decoupled from the
resonant light fields and trapped into a dark state, which depends
on the radiation amplitudes and phases. Perfectly phase correlated
laser fields, i.e. fields with matched Fourier
components, even though resonant, are not absorbed. In this paper we show, that in $%
\Lambda $-type excitation under the terms of EIT phase noise of
one laser field is transfered to the other one in a way, that
perfect phase correlation, i.e. $\omega _{2}-\omega _{1}=const.$,
is given for the two laser fields $\omega _{1}$ and $\omega _{2}$.
Such a perfect correlation is essential for high resolution in
quantum interference applications. Among the basic field
parameters' correlation processes in coherently prepared media
there is pulse matching \cite{harPRL70}, amplitude and phase
matching \cite{korPRA60}, matched photon statistics
\cite{agaPRL71}, and intensity- \cite{groOL22} and phase noise
correlation \cite{flePRL72,flePRA51}, which extends also to
quantized fields, including squeezing \cite{qcorrsqu} and quantum entanglement \cite%
{ent}. In experiments phase correlated laser waves are usually
produced via sideband modulation techniques (electrooptical-,
acoustooptical modulators, VCSELs \cite{affAPB70}) or optical
phase locking \cite{preAPB60}. Hence the accessible frequency
differences of phase correlated laser fields are restricted to the
electronically available frequency limits, presently of the order
of GHz. We show, that any pair of laser frequencies, even with
frequency spacing far beyond the GHz range, can be correlated in
phase in the EIT regime. It can be done easily with a simple
experimental setup, provided there is a suitable atomic or
molecular medium, whose energy level system allows to combine
resonantly the two frequencies in form of a $\Lambda $-type
excitation scheme.

\begin{figure}[htb]
\includegraphics[width=8cm]{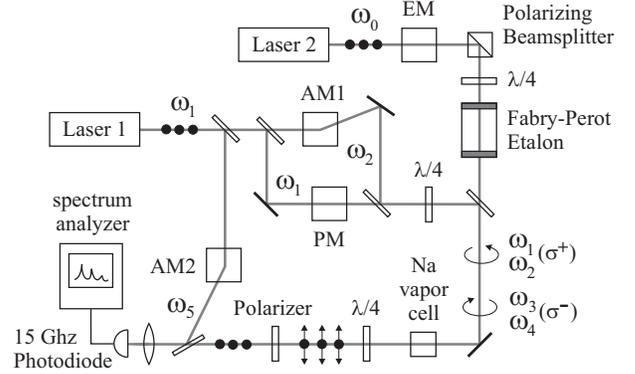}
\caption{Experimental setup: EM - electrooptical modulator, AM1/2
- acoustooptical modulator, PM - phase modulator.} \label{Fig.1}
\end{figure}

Experiments are performed on the $Na$ $D_{1}$ line (590nm) by use
of frequency
stabilized cw-dye lasers (Fig.\ref{Fig.1}). The two hyperfine sublevels $%
F=1,2$ of $Na$ ground state $3^{2}S_{1/2}$, spaced by $\omega _{12}=1771.6$
MHz, are connected to $3^{2}P_{1/2},F=2$ by $\sigma ^{+}$-polarized laser
beams (frequencies $\omega _{1},\omega _{2}$) and $\sigma ^{-}$-polarized
laser beams (frequencies $\omega _{3},\omega _{4}$), thus forming the $%
\Lambda $- and closed double-$\Lambda $ transition schemes. Laser
1 emitting frequency $\omega _{1}$ is stabilized to transition
$3^{2}S_{1/2},F=1$ - $3^{2}P_{1/2},F=2$. An acoustooptical
modulator (AM1) driven at 1771,6 MHz produces the first
negative order modulation sideband $\omega _{2}$, resonant to transition $%
3^{2}S_{1/2},F=2$ - $3^{2}P_{1/2},F=2$. An electrooptical phase
modulator (PM), which causes a phase shift of 16,3 mrad/V, is used
to modulate the phase of a part of $\omega _{1}$ by a 100 MHz band
limited white frequency noise. For investigations in double-$%
\Lambda $ configuration a second pair of frequency components ($\omega
_{3},\omega _{4}$) is generated by means of an electrooptical modulator (EM)
driven at 885,8 MHz. The two first order modulation sidebands of carrier $%
\omega _{0}$ (Laser 2 is stabilized to the $\Lambda $-crossover resonance $%
3^{2}S_{1/2},F=1,2$ - $3^{2}P_{1/2},F=2$) match the $Na$ ground
state hyperfine splitting. To avoid an additional absorption
background the carrier frequency $\omega _{0}$ is suppressed by an
electronically stabilized Fabry-Perot etalon. It is combined with
a polarizing beamsplitter and a quarter wave plate to circumvent
intense retroreflection into the dye laser.
A defined phase relation between frequency pairs ($%
\omega _{1},\omega _{2}$) and ($\omega _{3},\omega _{4}$) is
ensured by an internal frequency synchronization of the two
frequency generators driving AM1 and EM \cite{korPRA59}. The
frequency pairs are prepared in circular polarization and
transmitted collinearly through the absorption cell at typical
incident intensities of 200 mW/cm$^{2}$ per frequency component.
After the cell they are separated by a quarter wave plate and a
polarizer and observed separately with a 15 GHz InGaAs Schottky
photodiode using heterodyne spectroscopy. To observe the spectral
noise distribution $S_{i}(\tau ,\omega )$, i.e. the intensity
spectrum of phase noise of a single laser component with carrier frequency $%
\omega _{i}$, a part of $\omega _{1}$ is shifted by 260 MHz using
acoustooptical modulator AM2 ($\omega _{5}=\omega _{1}+260MHz$)
and superposed with the transmitted $\omega _{i}$ laser beam. As
$\omega _{5}$ is free of noise, the spectrum $S_{i}(\tau ,\omega
)$ can be observed directly by taking the beat signal $S_{i5}$ at frequency $%
\left\vert \omega _{i}-\omega _{5}\right\vert $ via a 2.8 GHz
electronic spectrum analyzer. The optical density $\tau (T)$ was
calibrated via an absorption measurement on transition $%
3^{2}S_{1/2},F=2$ - $3^{2}P_{1/2},F=2$. The vapor cell is a 1 cm$%
^{3}$ cube with a sidearm containing the sodium reservoir. Vapor
density (and $\tau $) is controlled via this reservoir
temperature, which is stabilized with an accuracy of 1${^{\circ
}}$C. The windows are kept at higher temperature to avoid
darkening. The cell is placed inside an arrangement of three
mutually orthogonal Helmholtz coils to compensate stray magnetic
fields.

A theoretical analysis of the correlation of phase fluctuations for EIT in $%
\Lambda $ systems is performed in refs. \cite{flePRL72,flePRA51}.
It is shown, that the spectrum $W_{\psi }$ of phase-difference
fluctuations $\delta \psi =\delta \varphi _{1}-\delta \varphi
_{2}$ of two laser fields $\omega _{1}$ and $\omega _{2}$ decays
with the propagation distance: $W_{\psi }\left( z,\omega \right)
=W_{\psi }\left( 0,\omega \right) \exp \left(
-\int_{0}^{z}2\kappa \left( z^{\prime },\omega \right) dz^{\prime }\right) $%
. Here $z$ is the propagation distance, and $%
\omega $ is the noise frequency. Contributions from the atomic
noise are neglected, which is justified
under the conditions of EIT. Obviously the laser phase fluctuations $%
\delta \varphi _{1}$ and $\delta \varphi _{2}$ become more and
more correlated
with the propagation distance. Slight extension of the theory in \cite%
{flePRL72,flePRA51}, assuming equal dipole moments, decay rates
and Rabi frequencies of the involved atomic transitions and no
phase correlation before interaction, gives for the individual
spectra of phase fluctuations $W_{11}$, $W_{22}$ and for the
cross-correlation $W_{12}$ the following approximate dependencies
on the propagation path
\begin{eqnarray}
\begin{split}
W_{11}(z,\omega )=& \frac{1}{4}\left( W_{11}(0,\omega )+W_{22}(0,\omega
)\right) \left( 1+e^{-2x}\right) + \\
& \frac{1}{2}\left( W_{11}(0,\omega )-W_{22}(0,\omega )\right)
\end{split}
\label{w1}\\
\begin{split}
W_{22}(z,\omega )=& \frac{1}{4}\left( W_{11}(0,\omega )+W_{22}(0,\omega
)\right) \left( 1+e^{-2x}\right) - \\
& \frac{1}{2}\left( W_{11}(0,\omega )-W_{22}(0,\omega )\right)
\end{split}
\label{w2}\\
\mathrm{Re}W_{12}(z,\omega )=\frac{1}{4}\left( W_{11}(0,\omega
)+W_{22}(0,\omega )\right) \left( 1-e^{-2x}\right)
\label{rw12}\\
\mathrm{Im}W_{12}(z,\omega )=\frac{1}{2}\left( W_{11}(0,\omega
)-W_{22}(0,\omega )\right) e^{-x}
\label{iw12}
\end{eqnarray}
with $x=\int_{0}^{z}\kappa \left( z^{\prime },\omega \right)
dz^{\prime }$. These eqs. show, that two laser fields transfer and
exchange their noise properties in the course of propagation.
After a sufficient long propagation path, the noise spectra of
both fields are identical and the fields are perfectly
correlated: $\left\vert 2W_{12}\right\vert =\sqrt{W_{11}W_{22}}$ \cite%
{mandel}. If initially only one of the fields has fluctuations
above shot noise ($W_{11}\left( 0,\omega \right) \neq 0$,
$W_{22}\left( 0,\omega \right)
=0$), then we observe from the above eqs. the noise transfer from component $%
\omega _{1}$ to the initially noise free component $\omega _{2}$:
$W_{22}\left( z,\omega \right) =\frac{1}{4}W_{11}\left( 0,\omega
\right) \left(
1-e^{-x}\right) ^{2}$. Simultaneously the cross-correlations $\mathrm{Re}%
W_{12}\left( z,\omega \right) $ grow exponentially at comparable rate.

\begin{figure}[htb]
\includegraphics[width=8cm]{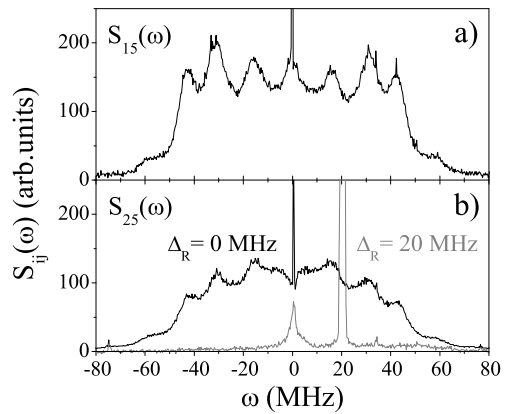}
\caption{Phase noise transfer at $\protect\tau=7.3$: Intensity
spectra $S_{15}\left( \protect\omega\right) $ (a) and $%
S_{25}\left( \protect\omega\right) $ (b), representing the spectra
of phase fluctuations $S_{1}\left(
\protect\tau,\protect\omega\right) $ and $S_{2}\left(
\protect\tau,\protect\omega\right) $ respectively; $\Delta_{R}=0$
(black curve) and $\Delta_{R}=20MHz$ (gray curve).} \label{Fig.2}
\end{figure}

In order to test this prediction experimentally, we observe the
intensity noise spectra $S_{1}\left( \tau ,\omega \right) $ and
$S_{2}\left( \tau ,\omega \right) $ by taking the beat signals
$S_{15}$ and $S_{25}$ respectively (the centers of the noise
spectra are normalized to zero frequency by subtracting the
frequency difference of the two carriers). Further we observe the
dependence of the FWHM $\Delta S_{12}$ of the beat signal $S_{12}
$ on $\tau $. An increasing optical path lengt $z$ is simulated by
changing the optical density $\tau $ via cell heating. Usually the
random frequency
jitter of light emitted by two different lasers is uncorrelated, and the width $%
\Delta S_{12}$ of the beat signal at frequency difference $\omega
_{2}-\omega _{1}$ is the sum of the linewidths of the two lasers. Phase
correlation due to EIT improves with $\tau $, thus $\Delta S_{12}(\tau )$,
similar to $W_{\psi }(z)$, reflects the degree of correlation between
different frequency sidebands of $\omega _{1}$ and $\omega _{2}$. $\Delta
S_{12}=0$ corresponds to perfect correlation. Fig.\ref{Fig.2}(a) shows a
measurement of the beat signal $S_{15}$, that represents the spectrum of
phase fluctuations $S_{1}(\tau ,\omega )$ modulated onto carrier $\omega _{1}
$.

\noindent

The wavelike structure is caused by the HF noise amplifier
characteristic, the central peak occurs at the exact carrier
frequency difference. The measurement of beat signal $S_{25}$
(Fig.\ref{Fig.2}(b) - black curve) exactly represents the spectrum
of phase fluctuations $S_{2}(\tau ,\omega )$: A noise transfer
from frequency component $\omega _{1}$ to the initially noise free
frequency component $\omega _{2}$, as predicted by eq. (\ref{w2}),
is obvious.

Under the same assumptions as for eqs. (\ref{w1})-(\ref{iw12}) we
derived the noise transfer rate
\begin{equation}
\kappa =\kappa _{0}\frac{\Gamma _{g}^{2}}{\Gamma _{g}^{2}+\left( \omega
-\Delta _{R}\right) ^{2}}\frac{\omega ^{2}}{\Gamma _{g}^{2}+\omega ^{2}},
\label{kappa-e}
\end{equation}%
where $\kappa _{0}=4\mu \left\vert \Omega \right\vert ^{2}N/\gamma
^{2}\Gamma _{g}$ is the maximum rate, with transition coupling
elements $\mu _{j}=\omega _{j}d_{j}^{2}/\hbar c$, dipole
transition
moments $d_{j}$, spontaneous decay rates $\gamma _{j}$, Rabi frequencies $%
\Omega _{j}=d_{j}E_{j}/\hbar $ (all assumed equal), atom density $%
N$, Raman detuning $\Delta _{R}=\omega _{1}-\omega _{2}-\omega
_{12}$, and the transparency window\ width $\Gamma _{g}=\Gamma
+2\left\vert \Omega \right\vert ^{2}/\gamma $. $\Gamma $ is the
dark state coherence decay rate. The noise transfer rate $\kappa $
shows a typical Lorentzian profile with respect to the Raman
detuning $\Delta _{R}$, with the width $\Gamma _{g}$. Only for a
small band of Raman detuning $\Delta _{R}$ around the noise
frequency $\omega $ the transfer rate is of considerable magnitude
- thus the efficient noise
transfer is due to EIT. This fact is supported by our measurements: In Fig.%
\ref{Fig.2}(b), the second (gray) curve is obtained for $\Delta
_{R}$ larger than the transparency window width. We see, that no
efficient noise transfer occurs, except for the beat frequencies
$\omega \approx \Delta _{R}=20MHz$, in correspondence with
(\ref{kappa-e}). We have performed such measurements for series of
different Raman detunings and observed, that the noise transfer
efficiency depends on $\Delta _{R}$ as a Lorentzian function, in
very good agreement with eq. (\ref{kappa-e}).

\begin{figure}[htb]
\includegraphics[width=8cm]{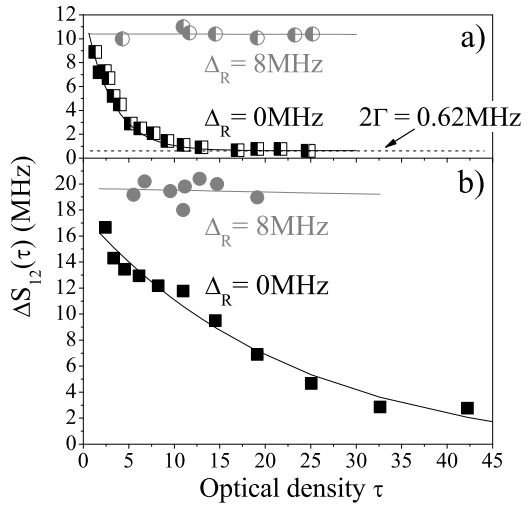}
\caption{FWHM $\Delta S_{12}\left( \protect\tau\right) $ of the $\protect%
\omega_{1}/\protect\omega_{2}$ beat signal in dependence on
$\protect\tau$ at zero- and 8MHz Raman detuning: (a) In the
$\Lambda$-scheme the exponential decay is limited by $\Gamma$; (b)
In the double-$\Lambda$-scheme the exponential decay
asymptotically approaches zero.} \label{Fig.3}
\end{figure}

An important feature follows from the factor $\omega ^{2}/\left( \Gamma
_{g}^{2}+\omega ^{2}\right) $ in eq. (\ref{kappa-e}): There is no
fluctuations correlation for frequencies inside the transparency window $%
\omega <\Gamma _{g}$, independent on the values of $\Delta _{R}$!
This is due to the adiabatic regime in this noise frequency range,
where small variations in the laser phase are so slow, that the
atom follows the evolution of the fields and remains in dark
state. Only high-frequency noise components $\omega
>\Gamma _{g}$ are correlated. As the laser intensity itself
exponentially decreases with optical density, and $\Gamma _{g}\sim
\Omega ^{2}$, the ultimate low-frequency threshold for phase
correlation is determined by $\Gamma $. This fact is clearly
demonstrated in our measurements: We observed the $S_{12}$ beat
signal at different optical densities $\tau $ for zero Raman
detuning. Since noise transfer happens, the spectra $S_{1}\left(
\protect\tau,\protect\omega\right) $ and $S_{2}\left(
\protect\tau,\protect\omega\right) $ are almost identical, and the
corresponding beat signal $\ S_{12}$ shows a Lorentzian profile
with a half width $\Delta S_{12}$ limited by $\Gamma _{g}$. The
FWHM $\Delta S_{12}\left( \tau \right) $ was evaluated and
depicted in Fig.\ref{Fig.3}(a): As $\omega _{1}$ and $\omega _{2}$
propagate, more and more lower noise frequencies $\omega $ become
correlated, and $\Delta S_{12}(\tau )$ exponentially decreases and
asymptotically approaches the limit set by $\Gamma $. The curve
$\Delta S_{12}\left( \tau \right) $ was fitted by an exponential
decay function yielding a dark state coherence decay rate $\Gamma
=0.3(1)MHz$, which is in good agreement with measurements of
$\Gamma $ \cite{husPRA63} made independently of the present
experiment. Additionally measurements of $\Delta S_{12}(\tau )$ at
8MHz Raman detuning show, that outside the EIT regime there is no
correlation effect at all!

\begin{figure}[htb]
\includegraphics[width=8cm]{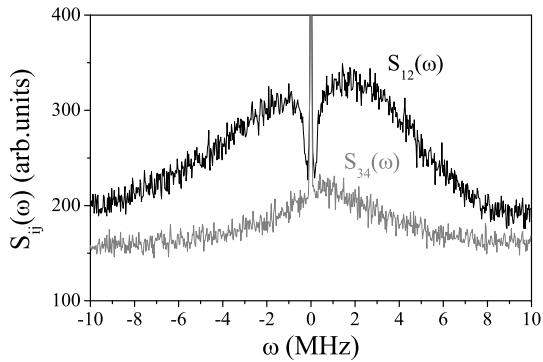}
\caption{ Intensity spectra $S_{12}\left( \protect\omega\right) $ and $%
S_{34}\left( \protect\omega\right) $, taken separately at $\protect\tau=7.3$
in case of four-frequency excitation in double-$\Lambda$-scheme.}
\label{Fig.4}
\end{figure}

In principle, the rate $\Gamma $ can be made very small if the low lying
states $\left\vert 1\right\rangle $ and $\left\vert 2\right\rangle $ of a $%
\Lambda $ system are close in energy. However, for a considerable energy
difference of $\omega _{1}$ and $\omega _{2}$ (e.g., in the optical range), $%
\Gamma $ will be quite large, of the order of $\gamma $. Moreover,
the correlation mechanism requires EIT, which in turn requires
sufficient high intensities $\left\vert \Omega \right\vert ^{2}\gg
\Gamma \gamma $, so that the transparency window gets even wider.
Thus, in such realistic cases, only small parts of the phase noise
spectrum can be correlated. This problem can be avoided if one
uses atom excitation in the double $\Lambda $ scheme: Here the
phase noise spectra have approximately the same dependence on the
propagation length as above - eqs. (\ref{w1})-(\ref{iw12}), but
similar relations are valid for any pair of the four participating
frequency components. Thus noise transfer and correlation of phase
fluctuations occur between all four radiation fields. Essential
for the double-$\Lambda $ system is the different noise transfer
coefficient, which for the stationary situation is given by
\begin{equation}
\kappa \simeq \kappa _{0}\frac{\Gamma _{g}^{2}\cos \varphi _{0}+\left(
\omega -\Delta _{R}\right) ^{2}\left( 1+\cos \varphi _{0}\right) }{\Gamma
_{g}^{2}+\left( \omega -\Delta _{R}\right) ^{2}}.  \label{kappad}
\end{equation}
Here $\kappa _{0}$ is the same as in eq. (\ref{kappa-e}), and
$\Gamma
_{g}=\Gamma +4\left\vert \Omega \right\vert ^{2}/\gamma $ for a double-$%
\Lambda $ system. $\varphi _{0}=(\varphi _{1}-\varphi
_{2})-(\varphi _{3}-\varphi _{4})$ is the value of the mean
relative phase of the transition excitation loop \cite{korPRA60}
at a given optical length. The phase $\varphi _{0}$ itself evolves
with the propagation distance, and inside the transparency window
$\left( \omega -\Delta _{R}\right) <\Gamma _{g}$ the phase
$\varphi _{0}$ rapidly (with the rate $\sim 4\kappa _{0}$)
approaches the value $2\pi n$ \cite{korPRA60}, while outside the
transparency window it changes very slowly, and on a scale $\kappa
_{0}^{-1}$ $\varphi _{0}$ is almost constant. For noise
frequencies inside the transparency window $\kappa \simeq \kappa
_{0}\cos \varphi _{0}$, while outside $\kappa \simeq \kappa
_{0}\left( 1+\cos \varphi _{0}\right) $. Consequently the noise
transfer coefficient is not zero in the whole frequency range. In
contrast to EIT in the $\Lambda $ system, the correlation of phase
fluctuations takes place for all noise frequencies $\omega $,
including the low-frequency range! This unlimited
correlation is demonstrated in the experiment: After transmission in double-$%
\Lambda $ excitation the beat signals $S_{12}(\omega )$ and
$S_{34}(\omega )$ (the latter is initially free of noise) are
observed separately (Fig.\ref{Fig.4}): Due to
the noise amplifier's cut-off frequency of 0.5 MHz the phase noise spectrum $%
S_{1}\left( \tau ,\omega \right) $ and accordingly the beat signal $%
S_{12}(\omega )$ shows a distinct dip around zero peak.

 Such a pronounced
dip is not found in the $S_{34}\left( \omega \right) $ intensity
spectrum, which confirms a phase noise transfer without frequency
limits for the double-$\Lambda $ regime. Analogous to the
measurements in Fig.\ref{Fig.3}(a)
we evaluated the FWHM $\Delta S_{12}\left( \tau \right) $ (Fig.\ref%
{Fig.3}(b)). In accordance with our model $\Delta S_{12}\left( \tau \right) $
exponentially decreases and (within the margin of fitting error)
asymptotically approaches zero.

In double-$\Lambda $ excitation the noise transfer coefficient
$\kappa \neq 0$ almost independent on the laser intensity, also
for large $\Gamma $ (e.g. for large carrier frequency
differences). Inside the transparency window the correlation
happens, in general, slower than in the $\Lambda $ system, as can
be seen from the different decay rates of $\Delta S_{12}\left(
\tau \right) $ in Fig.\ref{Fig.3}. This is the price to pay for
low-frequency noise correlation. It is interesting that for a
singular point $\varphi _{0}=\pi $, where no atomic coherence is
built up, the transfer rate is negative for all noise frequencies
and at all propagation distances (no phase matching occurs at
$\varphi _{0}=\pi $, see ref \cite{korPRA60}): The noise grows.
This fact has also been mentioned in ref. \cite{flePRA51}.

The double-$\Lambda $ system can be applied for a phase
correlation of lasers with substantially different wavelengths
(e.g. correlation of UV with IR) up to the shot noise (and even
beyond - using entanglement \cite{gloQP}). In practice a
double-$\Lambda $ excitation scheme is established easily - each
of the two fields of an appropriate $\Lambda $ system can be
shifted in frequency by an equal amount using sideband modulation,
and afterwards all four resulting fields are superimposed in the
medium. The process also works in the degenerated double-$\Lambda
$ configuration \cite{maiEPL31}. Here the mean relative phase
$\varphi _{0}$ is constant, and can easily be controlled and put
to zero \cite{husJMO49}. Such a setup might be relevant for any
EIT application to modern nonlinear optics, where standard phase
correlation techniques do not suffice. Besides high precision
spectroscopy we expect promising applications in quantum
information processing \cite{lukPRL84}, quantum
state engineering \cite{patQP} or long distance quantum communication \cite%
{duaNAT414}. As to the realization of a quantum repeater, quantum
correlated photon pairs have already been generated
\cite{photpairs} using the EIT-based technology of light pulse
storage \cite{storage}. Phase correlated excitation of optical
materials with high nonlinearities and low loss might well become
essential for the controlled generation of entangled states and
quantum logic operations in future quantum computer design.

\textit{We want to thank I. E. Mazets and M. Fleischhauer for very useful
discussions. This work was supported by the Austrian Science Foundation
under project number P 12894.}





\end{document}